\documentclass[journal]{IEEEtran}
\usepackage{amsmath,amsfonts}
\usepackage{mathrsfs}
\usepackage{algorithmic}
\usepackage{algorithm}
\usepackage{array}
\usepackage{textcomp}
\usepackage{stfloats}
\usepackage{url}
\usepackage{hyperref}
\hypersetup{
	colorlinks=true,
	linkcolor=blue,
	filecolor=blue,
	urlcolor=blue,
	citecolor=blue,
}
\usepackage{xcolor}
\usepackage{flushend}
\usepackage{verbatim}
\usepackage{graphicx}
\graphicspath{{Figures/}}
\usepackage{epstopdf}
\usepackage{cite}

\usepackage{multirow}
\usepackage{bm}
\usepackage{siunitx}
\usepackage{booktabs}
\usepackage{tabularx}

\usepackage{amssymb}
\def\BibTeX{{\rm B\kern-.05em{\sc i\kern-.025em b}\kern-.08em T\kern-.1667em\lower.7ex\hbox{E}\kern-.125emX}}

\begin{document}

	\title{Implicit-Reasoning-Enabled Elastic Semantic Communications for AI-Agent Networks}
\author{Kaiwen Yu,~\IEEEmembership{Member,~IEEE},
	Gang Wu,~\IEEEmembership{Senior Member,~IEEE}, Jiancun Fan,~\IEEEmembership{Senior Member,~IEEE}, and Saviour Zammit,~\IEEEmembership{Senior Member,~IEEE}
	
	\thanks{This work was supported in part by the National Key R\&D Program of China under Grant 2026YFE0104900; in part by GEMSS-6G, which received funding from Xjenza Malta (XM) and the Ministry of Science and Technology of the People’s Republic of China (MOST) through the SINO-MALTA Fund 2025 Call (Science and Technology Cooperation); in part by the National Natural Science Foundation of China under Grant 62601131; and in part by the Natural Science Foundation of Sichuan Province under Grant 2026NSFSC1423.}
	
	\thanks{Kaiwen Yu and Gang Wu are with the National Key Laboratory of Wireless Communications, University of Electronic Science and Technology of China, Chengdu 611731, China. Gang Wu is also with the Tianfu Jiangxi Laboratory, Chengdu, China (e-mail: \{yukaiwen; wugang99\}@uestc.edu.cn).}
	
	\thanks{Jiancun Fan is with the School of Information and Communications Engineering, Xi’an Jiaotong University, Xi’an, Shaanxi 710049, China (e-mail: fanjc0114@gmail.com).}
			
	\thanks{Saviour Zammit is with the Department of Communications and Computer Engineering, University of Malta, MSD 2080 Msida, Malta (e-mail: saviour.zammit@um.edu.mt).}
}

\maketitle

\begin{abstract}
	AI-agent networks introduce new communication demands driven by proactive interactions and autonomous task execution. Existing adaptive semantic communication schemes primarily adjust semantic representations and resource configurations within predefined processing scales and capability sets, which may be insufficient for evolving task requirements. To address this gap, we introduce elastic semantic communications (ESC), a task-driven framework that extends semantic adaptation to the runtime execution and participation of task-supporting capabilities. ESC comprises two complementary dimensions: vertical elasticity adjusts the execution intensity of existing capabilities, while horizontal elasticity reconfigures the participating capability set. The framework connects these dimensions through the expected marginal value of additional information and computation. To realize ESC, we propose an implicit-reasoning-enabled device--edge mechanism that supports variable-depth local inference through shared-parameter state updates and edge-side continuation through wireless transfer of intermediate reasoning states. Experiments on Sudoku-Extreme demonstrate that the benefits of additional inference depend on wireless conditions, while task-aware edge invocation can preserve task accuracy with substantially lower state-transfer overhead.
\end{abstract}

\section{Introduction}
\label{sec:introduction}

\IEEEPARstart{T}{he} growing use of AI agents in distributed applications is creating new demands on wireless communication. Such agents collaborate through sensing, communication, inference, and decision-making to accomplish shared tasks~\cite{zhang2026evolvingintelligentcomplexsystems,chen2025internet}. In these networks, information exchange primarily supports downstream inference and decision-making rather than merely reproducing source data at the receiver. Semantic communication, which emphasizes task-relevant information, therefore offers a promising paradigm for AI-agent networks.

Unlike many conventional communication services that operate with predefined service requirements and configurations, AI agents act as autonomous and proactive task executors. They can initiate communication, revise task plans, and request additional information or computing resources as tasks evolve~\cite{yu2025semanticdrivenaiagentcommunications}. Consequently, the information required for task execution, the amount of computation needed, and the capabilities involved may all change dynamically at runtime.

Existing adaptive semantic communication schemes address varying operating conditions by adjusting semantic representations, transmission strategies, and communication--computation resource allocation~\cite{yu2025distributed,zhang2022deep}. While these mechanisms remain essential, autonomous task execution also requires runtime decisions on the execution intensity of existing capabilities and the participation of additional ones. This calls for extending semantic adaptation beyond transmission and resource configurations to the runtime execution and participation of task-supporting capabilities.

Recent advances have explored complementary aspects of implicit reasoning and agent networking. At the model level, implicit reasoning models perform inference through iterative latent-state updates with shared parameters, supporting adjustable inference depth at test time~\cite{huang2026equilibrium}. In wireless semantic communication, the relationship between effective computation depth and channel conditions has also been investigated~\cite{yu2026effective}. Meanwhile, research on the Internet of Agents has examined architectures for large-scale agent networking and inter-agent coordination~\cite{zheng2026networking}. However, additional computation does not necessarily translate into better task performance, particularly when the available semantic information is degraded. Similarly, acquiring additional information is beneficial only when it contributes to downstream task decisions. Leveraging external computing resources also requires wireless information exchange, introducing communication costs and uncertainty. The resulting challenge is to determine when additional information or computation is worthwhile under evolving task requirements and wireless conditions.

Motivated by this challenge, we introduce elastic semantic communications (ESC), a capability-centric framework that extends semantic adaptation to the runtime execution and participation of task-supporting capabilities. ESC encompasses two complementary dimensions: vertical elasticity adjusts the execution intensity of existing capabilities, while horizontal elasticity reconfigures the participating capability set. Task-driven orchestration relates these decisions to the expected marginal value of information and computation, considering task feedback and wireless operating conditions. As a concrete realization, we develop an implicit-reasoning-enabled device--edge mechanism in which shared-parameter state updates support variable-depth local inference, while intermediate reasoning states can be transferred to compatible edge computing resources for continued processing without restarting inference. In this realization, a task- and channel-aware policy estimates the expected task-performance gain of edge continuation from observable local inference statistics and the state-transfer SNR. Experiments on Sudoku-Extreme demonstrate that the benefits of additional inference depend on wireless conditions, while task-aware edge invocation can maintain task accuracy with substantially lower state-transfer overhead.


\begin{figure*}[!t]
	\centering
	\includegraphics[width=6.4in]{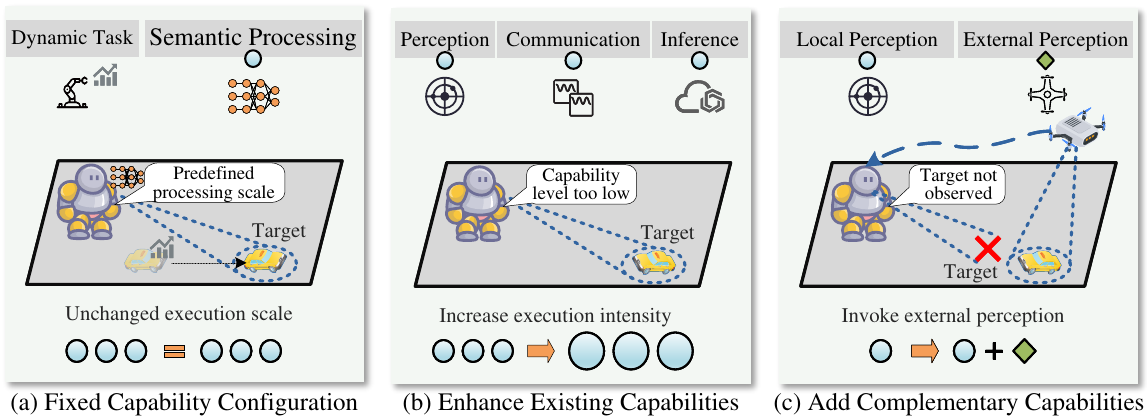}
	\caption{Illustrative capability responses to changing task requirements: (a) fixed processing configuration, (b) enhancing existing capabilities through adjustable processing effort, and (c) adding complementary capabilities to obtain missing task information.}
	\label{Fig1_ESC_Motivating_Scenarios}
\end{figure*}

%
%
%
%

\section{From Adaptive to Elastic Semantic Communications}
\label{sec:requirements}

Adaptive semantic communication has evolved across multiple levels, from channel-aware transceiver operation to task-guided information acquisition and network resource optimization. To cope with varying wireless conditions, two-way semantic communication exploits channel reciprocity and learned channel distributions to enable local transceiver adaptation without exchanging training feedback~\cite{yu2024twsc}. Partial sampling-based semantic communication uses receiver feedback to adaptively select task-relevant image regions over successive sampling and transmission rounds~\cite{yu2025partial}. Channel-aware vision-token compression further enables progressive rate adaptation of intermediate representations in partitioned vision-language models~\cite{hua2026channelaware}. At the network level, distributed hierarchical deep reinforcement learning coordinates beamforming and power allocation on a longer timescale and semantic compression rates on a shorter one to accommodate changing channel conditions and semantic requirements~\cite{yu2025distributed}. Semantic-network orchestration has also explored knowledge-base management and the coordination of network and computing resources~\cite{shokrnezhad2024semantic}.

These advances demonstrate that semantic adaptation can respond to diverse operating conditions across transmission, information acquisition, and resource management. However, adapting existing mechanisms does not necessarily address every requirement emerging during autonomous task execution. Fig.~\ref{Fig1_ESC_Motivating_Scenarios} illustrates this distinction. In Fig.~\ref{Fig1_ESC_Motivating_Scenarios}(a), an agent processes target observations with a predefined processing budget. As task demands change, the execution scale remains fixed, potentially resulting in insufficient processing for demanding tasks or unnecessary computation for simpler ones. When the required information is available but processing is insufficient, increasing the execution effort of existing capabilities may improve task performance, as illustrated in Fig.~\ref{Fig1_ESC_Motivating_Scenarios}(b). In contrast, Fig.~\ref{Fig1_ESC_Motivating_Scenarios}(c) depicts a target that cannot be observed locally. In this case, additional processing cannot by itself provide the missing observation, and an external sensing capability may be required. These cases highlight the importance of distinguishing insufficient processing from missing task information, rather than treating every unmet requirement as a need for more computation.

This distinction motivates a broader scope of runtime adaptation. Beyond adjusting transmission and resource configurations, evolving tasks may require changes to the capabilities involved in execution and the information exchanged among them. Introducing an external capability can provide new observations, specialized functions, or additional computing resources, but also changes the communication and coordination required to complete the task. Such participation should therefore be determined by its expected contribution to task execution under the prevailing wireless and resource conditions. Elastic semantic communications builds on existing adaptive mechanisms by extending runtime decisions to the organization of task-supporting capabilities, as detailed in the following section.

\section{Task-Driven Elastic Semantic Communications Framework}
\label{sec:framework}

\begin{figure*}[!t]
	\centering
	\includegraphics[width=6.5in]{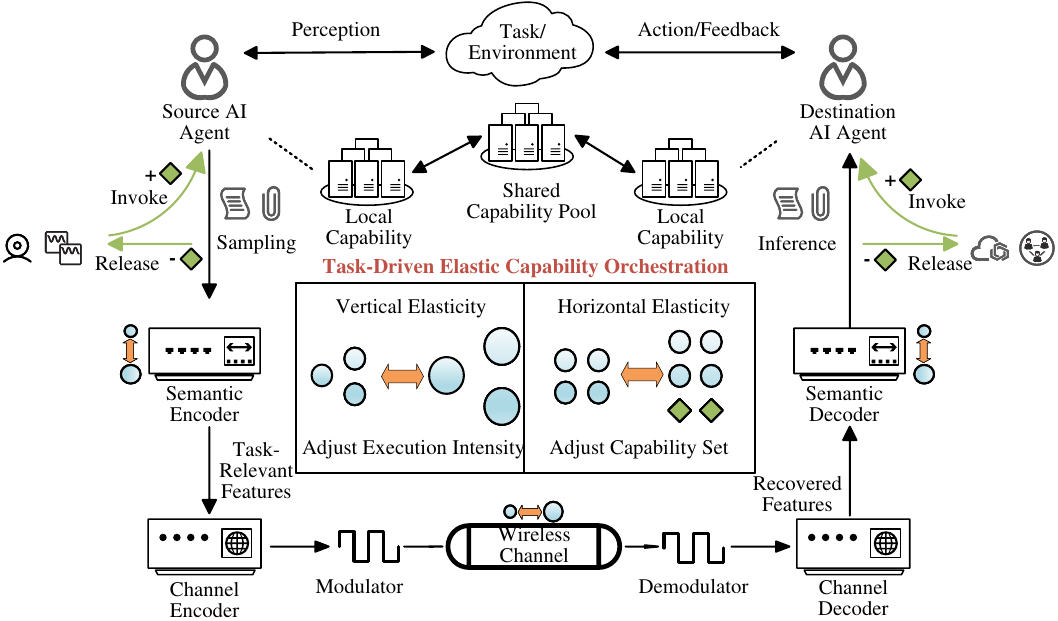}
	\caption{Task-driven ESC framework for wireless AI-agent networks. Vertical elasticity adjusts the execution effort of participating capability instances, while horizontal elasticity reconfigures the capability set through invocation or release. Semantic communication supports task-relevant information exchange across agents, enabling capability orchestration guided by the expected value of additional information and computation.}
	\label{Fig2_ESC_Framework}
\end{figure*}

From a wireless semantic communication perspective, AI-agent tasks rely on three interdependent processes: acquiring task-relevant information, exchanging information across nodes, and performing task inference and execution~\cite{yu2025semanticdrivenaiagentcommunications}. ESC organizes the capabilities supporting these processes according to evolving task requirements. As illustrated in Fig.~\ref{Fig2_ESC_Framework}, a capability denotes a task-supporting function, such as sensing, communication, or inference, while a capability instance refers to its executable deployment on an agent or computing node.

\subsection{Vertical Elasticity: Scaling Existing Capabilities}

Vertical elasticity adjusts the execution effort of participating capability instances without changing the participating set. Its control variables include sensing rounds, the amount or frequency of semantic information exchange, and processing depth or reasoning iterations.
These adjustments affect task performance through coupled forms of value. Additional sensing provides observations whose value of information (VoI) depends on their expected contribution to downstream task decisions~\cite{russell1991metareasoning}. Further processing exploits the acquired representations, with its value of computation (VoC) reflecting the expected improvement from additional refinement~\cite{russell1991metareasoning}. Communication adaptation determines how effectively task-relevant semantics are delivered over wireless channels, thereby bounding the VoI preserved at the receiver and the VoC achievable through subsequent inference. For example, feedback-guided partial sampling adapts information acquisition according to task needs~\cite{yu2025partial}, while iterative reasoning scales computation according to the quality of received information.

\subsection{Horizontal Elasticity: Capability Reconfiguration}

Horizontal elasticity reconfigures the capability instances or functions participating in task execution, thereby expanding the information and computation options accessible to the task. This includes introducing compatible execution instances to extend computing resources and invoking complementary capabilities to provide observations or functionality absent in the current configuration.
As illustrated in Fig.~\ref{Fig2_ESC_Framework}, external sensing broadens the observation space, whose VoI depends on its relevance and novelty relative to existing local knowledge. In contrast, invoking a compatible edge computing instance extends the processing resources dedicated to an ongoing task, with its VoC determined by whether continued computation can further refine the current estimate. Specialized models also introduce new functional pathways, enabling the agent to process heterogeneous data sources or apply alternative reasoning mechanisms.

Capability reconfiguration also alters the information dependencies among participating nodes. External capabilities require compatible task inputs and must return outputs usable by downstream components. The net value realized from external VoI or VoC therefore depends on the reliability and cost of the required wireless exchange. Horizontal elasticity thus reshapes both the participating capability set and the underlying semantic information flow.

\subsection{Task-Driven Elastic Capability Orchestration}

Task-driven orchestration uses the expected marginal VoI and VoC as a unified basis for elastic execution decisions. Based on observable task feedback and wireless conditions, it evaluates the potential net benefits of further processing local information, acquiring additional observations, or invoking external capabilities. These values are assessed relative to the current task state, reflecting the expected marginal improvement from each action rather than the static capacity of a capability.

The orchestration process weighs alternative actions, including returning the current output, scaling the execution effort of active capabilities, and reconfiguring the capability set. Their selection accounts for expected task gains, wireless transmission reliability, computation and communication costs, and capability compatibility. Vertical and horizontal decisions are interdependent but not necessarily sequential: new observations can unlock higher VoC for subsequent computation, while local processing feedback can reveal whether external assistance is warranted. As tasks evolve, updated execution feedback enables the relative value of these actions to be reassessed dynamically.

\section{Implicit-Reasoning-Enabled Elastic Execution}

As a concrete realization of ESC, we consider an implicit-reasoning-enabled device--edge system that supports variable-depth local inference and cross-node reasoning continuation through shared-parameter state updates.

\subsection{Implicit Reasoning for Elastic Execution}

\begin{figure*}[!t]
	\centering
	\includegraphics[width=6.5in]{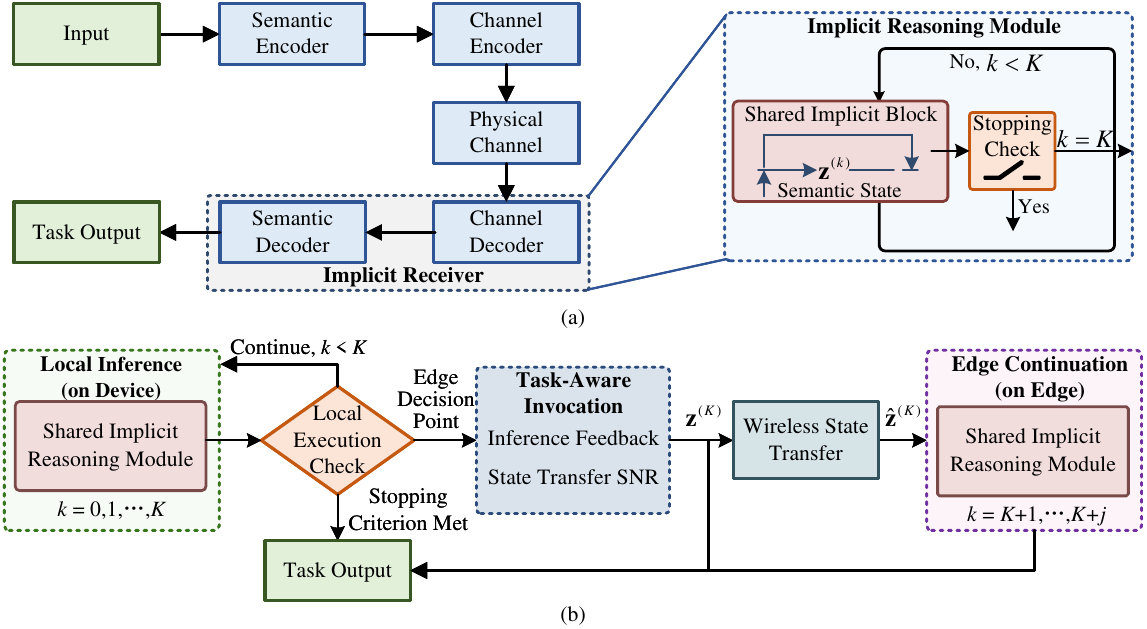}
	\caption{Implicit-reasoning-enabled elastic semantic execution. (a) Semantic communication system and receiver-side implicit reasoning with variable-depth state updates. (b) Task-driven execution involving local refinement and selective edge continuation. Intermediate reasoning states and task context enable cross-node continuation without restarting inference.}
	\label{Fig4_ESC_Realization}
\end{figure*}

Fig.~\ref{Fig4_ESC_Realization}(a) illustrates the integration of implicit reasoning into the receiver of a semantic communication system. Given the received semantic input $\mathbf{x}$, an implicit reasoning model defines a latent equilibrium $\mathbf{z}^\star$ satisfying
$
\mathbf{z}^\star=F_\theta(\mathbf{z}^\star;\mathbf{x}),
$
which is approached through iterative state updates
$
\mathbf{z}^{(k+1)}=F_\theta(\mathbf{z}^{(k)};\mathbf{x})
$
with shared model parameters $\theta$~\cite{huang2026equilibrium}. Starting from an initial latent state, the receiver repeatedly applies the same update function to refine its internal representation and produce a task output from the current reasoning state. Since the model parameters are shared across iterations, the executed inference depth can be adjusted without changing the deployed model architecture or parameters.

A further advantage of this iterative formulation is its ability to support reasoning continuation across computing nodes. As illustrated in Fig.~\ref{Fig4_ESC_Realization}(b), the device can perform an initial set of state updates and transmit the resulting intermediate state, together with the required task context, to a compatible edge model. The edge model then resumes the iterative reasoning process from the received state rather than restarting from the initial state. This enables additional computing resources to contribute to an ongoing inference process while preserving the progress represented by the transferred state. However, wireless transmission may perturb the intermediate state and affect the quality of subsequent reasoning.

\subsection{Task- and Channel-Aware Edge Continuation}

The ability to continue reasoning at the edge does not imply that additional computation always improves task performance. Numerical convergence and task correctness are not equivalent: a stable reasoning state may still produce an incorrect result when the available semantic information is insufficient, while a correct task output may emerge before numerical convergence. Moreover, transferring intermediate states introduces wireless transmission uncertainty and communication overhead. Edge assistance should therefore be selected according to its expected contribution to task performance rather than solely on local computation limits or convergence behavior.

To enable task- and channel-aware edge invocation, a lightweight continuation-gain predictor, such as a regression tree, can be trained offline using inference statistics collected from repeated local executions and their corresponding edge-continuation outcomes. These data are used to fit the relationship between observable inference feedback and the expected benefit of additional reasoning under different wireless conditions. Once calibrated, the predictors are deployed online, where the state-transfer signal-to-noise ratio (SNR) selects the corresponding model to estimate the continuation gain for the current task. Edge assistance is invoked only when the predicted gain is positive; otherwise, the device returns its local output.

\section{Performance Evaluation}
\label{sec:example}

We evaluate the proposed realization on the Sudoku-Extreme benchmark using a pretrained equilibrium model~\cite{huang2026equilibrium}, which remains frozen throughout the experiments. A puzzle is considered correctly solved only when all 81 predicted cells match the reference solution. The wireless system consists of two communication links: an input link that delivers task representations to the reasoning agent and a state-transfer link that enables edge-side inference continuation. Both links are modeled as complex additive white Gaussian noise (AWGN) channels.

\subsection{Vertical Elasticity: Depth Scaling}

To evaluate inference-depth scaling, we use 1,024 test puzzles under varying input-channel conditions. Task representations are transmitted over the wireless channel with a fixed transmission length. The input-channel SNR varies from $-5$ to 30~dB, while the inference depth is set to $K\in\{1,2,4,8,16,32,64\}$. This setting allows us to examine the computation--accuracy tradeoff under varying wireless conditions.

\begin{figure}[!t]
	\centering
	\includegraphics[width=3.5in]{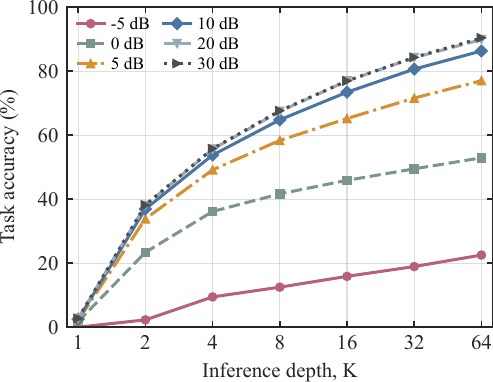}
	\caption{Task accuracy versus inference depth under different input-channel SNRs on the Sudoku-Extreme benchmark.}
	\label{Fig4_Elastic_Inference_Depth}
\end{figure}

Fig.~\ref{Fig4_Elastic_Inference_Depth} illustrates the relationship between inference depth, wireless channel quality, and task performance. Increasing inference depth generally improves task accuracy by providing additional reasoning computation. At 5~dB, increasing the depth from 8 to 32 iterations improves accuracy from 58.37\% to 71.61\%, while at 10~dB, the corresponding accuracy increases from 64.81\% to 80.66\%. However, the benefit of additional computation depends on the quality of the received semantic information. At $-5$~dB, task accuracy reaches only 22.55\% even after 64 iterations, compared with 86.33\% at 10~dB. These results demonstrate that inference depth and wireless channel conditions jointly determine task performance, motivating runtime adjustment of processing effort according to the available semantic information.

\subsection{Horizontal Elasticity: Task-Driven Edge Continuation}

For edge-continuation evaluation, the input-channel SNR is fixed at 10~dB, while the state-transfer SNR varies from $-5$ to 30~dB. The device performs eight local iterations before making an edge-invocation decision. When selected, the edge model continues inference for 24 additional iterations. The intermediate reasoning state and task context are transmitted over the wireless channel using uncoded analog mapping, with a fixed payload size for each invocation.
We compare device-only execution, always-edge continuation, and two calibrated invocation policies. The SNR-aware policy uses the average continuation gain at each state-transfer SNR, while the task-and-SNR-aware policy employs an SNR-specific depth-two regression tree using some candidate inference statistics (e.g., prediction confidence, entropy, and logit margins) extracted after eight local iterations. The predictors are calibrated on 256 puzzles and evaluated on 768 held-out puzzles.
The reported task accuracy is averaged over multiple runs with different random seeds and channel noise realizations.

\begin{figure}[!t]
	\centering
	\includegraphics[width=3.5in]{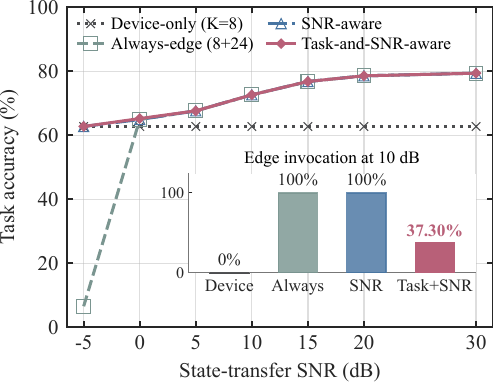}
	\caption{Task accuracy versus state-transfer SNR under selective edge continuation; the inset shows edge-invocation ratios.}
	\label{Fig5_end_edge_elastic_inference}
\end{figure}

Fig.~\ref{Fig5_end_edge_elastic_inference} illustrates how wireless state-transfer quality affects edge-assisted inference. With eight local iterations, device-only execution achieves 62.73\% task accuracy. Always-edge continuation reduces accuracy to 6.53\% at $-5$~dB due to unreliable state transfer, but improves it to 72.62\% at 10~dB. These results highlight the strong dependence of continuation gains on wireless channel conditions.
The SNR-aware policy avoids unfavorable state transfers under poor channel conditions but cannot distinguish tasks experiencing the same SNR. Consequently, it invokes the edge for all tasks at 10~dB. In contrast, the task-and-SNR-aware policy exploits observable inference statistics to selectively invoke edge continuation. At 10~dB, it achieves the same 72.62\% accuracy as always-edge execution with only 37.30\% edge invocations, reducing additional state-transfer overhead by 62.70\%.

\section{Open Challenges and Future Directions}
\label{sec:Directions}

The preceding evaluation illustrates how implicit reasoning supports elastic execution through inference-depth scaling and wireless edge continuation. Extending these mechanisms to practical AI-agent networks presents several open challenges.

\subsection{Defining and Measuring Semantic Sufficiency}

A fundamental challenge is determining whether the available semantic information and computation are sufficient for reliable task execution. Numerical convergence does not necessarily guarantee task correctness, while the correct outcome is generally unknown at runtime. Future research should develop task-aware sufficiency criteria based on observable indicators, such as prediction confidence and output stability, to support reliable stopping and continuation decisions under varying wireless conditions.

\subsection{Joint Orchestration of Vertical and Horizontal Elasticity}

Joint orchestration becomes more challenging as autonomous agents interact with heterogeneous capabilities under dynamic task and network conditions. Adjusting the execution intensity of existing capabilities may change the need for external assistance, while invoking additional capabilities introduces communication and coordination costs. Future mechanisms should coordinate these coupled decisions across different timescales, considering task performance, resource availability, and wireless conditions.

\subsection{Capability Discovery and Semantic Interoperability}

Horizontal elasticity requires discovering and selecting external capabilities that can contribute to ongoing tasks~\cite{guo2026agent}. Beyond resource availability, agents must understand the functions, interfaces, and semantic compatibility of candidate capabilities. Developing efficient capability discovery, semantic alignment, and task-level information integration mechanisms is essential for collaboration among heterogeneous AI agents.

\subsection{Elastic Semantic Interfaces and Signaling}

Practical ESC requires communication interfaces that support task-relevant information exchange and dynamic capability coordination~\cite{ren2026toward}. Transmitting semantic features, intermediate states, or task context introduces wireless overhead and reliability challenges. Task-aware state compression, robust joint source--channel coding, and lightweight control signaling are promising directions for enabling efficient and reliable elastic execution.

\section{Conclusion} 

This article introduced elastic semantic communications as a capability-centric framework for coordinating the execution and participation of task-supporting capabilities in wireless AI-agent networks. By considering the expected marginal value of information and computation, ESC provides a unified perspective for task-driven capability orchestration under evolving task and wireless conditions. We developed an implicit-reasoning-enabled realization that combines variable-depth local inference with wireless edge continuation and task-aware capability invocation. Experiments on Sudoku-Extreme demonstrated that the benefits of additional inference depend on wireless conditions and task-specific execution states. The evaluated invocation policy reduced wireless state exchanges while retaining the accuracy of always-edge execution. These results illustrate the potential of value-driven elastic execution for wireless AI-agent networks.

\bibliographystyle{IEEENetworkComSoc}
\bibliography{ESI_references}

\end{document}